\documentclass[superscriptaddress,twocolumn,showpacs,pre,aps]{revtex4-1}

\usepackage{graphicx}
\usepackage{amsmath, amsthm, amssymb, bm}

\begin{document}

\title{Fingerprinting molecular relaxation in deformed polymers}
\author{Zhe Wang}
\email{zwang.thu@gmail.com}
\affiliation{Biology and Soft Matter Division, Oak Ridge National Laboratory, Oak Ridge, Tennessee 37831, USA}
\author{Christopher N. Lam}
\affiliation{Center for Nanophase Materials Sciences, Oak Ridge National Laboratory, Oak Ridge, Tennessee 37831, USA}
\author{Wei-Ren Chen}
\affiliation{Biology and Soft Matter Division, Oak Ridge National Laboratory, Oak Ridge, Tennessee 37831, USA}
\author{Weiyu Wang}
\affiliation{Center for Nanophase Materials Sciences, Oak Ridge National Laboratory, Oak Ridge, Tennessee 37831, USA}
\author{Jianning Liu}
\affiliation{Department of Polymer Science, University of Akron, Akron, Ohio 44325, USA}
\author{Yun Liu}
\affiliation{Center for Neutron Research, National Institute of Standards and Technology, Gaithersburg, Maryland 20899, USA}
\author{Lionel Porcar}
\affiliation{Institut Laue-Langevin, B.P. 156, F-38042 Grenoble CEDEX 9, France}
\author{Christopher B. Stanley}
\affiliation{Biology and Soft Matter Division, Oak Ridge National Laboratory, Oak Ridge, Tennessee 37831, USA}
\author{Zhichen Zhao}
\affiliation{Department of Polymer Science, University of Akron, Akron, Ohio 44325, USA}
\author{Kunlun Hong}
\affiliation{Center for Nanophase Materials Sciences, Oak Ridge National Laboratory, Oak Ridge, Tennessee 37831, USA}
\author{Yangyang Wang}
\email{wangy@ornl.gov}
\affiliation{Center for Nanophase Materials Sciences, Oak Ridge National Laboratory, Oak Ridge, Tennessee 37831, USA}

\begin{abstract}
The flow and deformation of macromolecules is ubiquitous in nature and industry, and an understanding of this phenomenon at both macroscopic and microscopic length scales is of fundamental and practical importance. Here we present the formulation of a general mathematical framework, which could be used to extract, from scattering experiments, the molecular relaxation of deformed polymers. By combining and modestly extending several key conceptual ingredients in the literature, we show how the anisotropic single-chain structure factor can be decomposed by spherical harmonics and experimentally reconstructed from its cross sections on the scattering planes. The resulting wavenumber-dependent expansion coefficients constitute a characteristic fingerprint of the macromolecular deformation, permitting detailed examinations of polymer dynamics at the microscopic level. We apply this approach to survey a long-standing problem in polymer physics regarding the molecular relaxation in entangled polymers after a large step deformation. The classical tube theory of Doi and Edwards predicts a fast chain retraction process immediately after the deformation, followed by a slow orientation relaxation through the reptation mechanism. This chain retraction hypothesis, which is the keystone of the tube theory for macromolecular flow and deformation, was critically examined by analyzing the fine features of the two-dimensional anisotropic spectra from small-angle neutron scattering by entangled polystyrenes. It is shown that the unique scattering patterns associated with the chain retraction mechanism were not experimentally observed. This result calls for a fundamental revision of the current theoretical picture for nonlinear rheological behavior of entangled polymeric liquids.
\end{abstract}

\date{\today}

\pacs{83.10.Kn, 83.85.Hf, 83.80.Sg, 83.60.Df}
\maketitle

\section{Introduction}
The entanglement phenomenon is one of the most important and fascinating characteristics of long flexible chains in the liquid state \cite{Porter, Graessley, Lodge}. Our current understanding of the dynamics of entangled polymers is built upon the tube theoretical approach pioneered by de Gennes \cite{deGennes1}, Doi, and Edwards \cite{DE1, DE2, DE3, DE4, DE}. In his 1971 JCP paper \cite{deGennes1}, de Gennes demonstrated how the diffusion problem of a flexible chain could be understood in terms of reptation in the presence of fixed obstacles. A few years later, in a series of seminal publications \cite{DE2, DE3, DE4}, Doi and Edwards illustrated how the molecular motion under flow and deformation could be explained with the aid of the tube concept. The advent of the tube model has revolutionized the field of polymer dynamics and the predictions of the model about both the linear and nonlinear viscoelastic properties of entangled polymers have been significantly improved over the years, by incorporating additional molecular mechanisms such as contour length fluctuation \cite{Doi, MMc, LMc}, constraint release \cite{Daoud, Klein, Marrucci, MLD, MMcL}, and chain stretching \cite{MG, Pearson, Graham}. However, despite the remarkable success of the tube approach, particularly in the linear response regime, one of the central hypotheses of the model has so far eluded experimental confirmation.

In an effort to account for the \textit{nonlinear} rheological behavior, Doi and Edwards \cite{DE2} proposed a unique microscopic deformation mechanism for entangled polymers, which asserts that the external deformation acts on the tube, instead of the polymer chain \cite{Larson}. The \textit{chain retraction} within the affinely deformed tube would lead to non-affine evolution of chain conformation beyond the Rouse time, with entanglement strands being oriented but hardly stretched. This hypothesis, being a keystone of the tube model, stands in stark contrast to the elastic deformation mechanisms of many other alternative theoretical approaches such as the transient network model \cite{GT, ASLodge, Yamamoto1} where the affine deformation mechanism is adopted. While scattering techniques, particularly small-angle neutron scattering (SANS), have long been envisioned as the ideal tool for critical examination of this key hypothesis of the tube model, the SANS investigations in the past several decades have not led to a clear conclusion, with many questioning the validity of the non-affine mechanism \cite{Maconnachie, Boue1, Boue2, Mortensen1, Mortensen2, Schroeder}, some claiming support \cite{Bent, Blanchard, Graham2}, and others being silent on this issue \cite{Muller, Hassager}. Moreover, recent experimental studies \cite{SQWangJCP, SQWangMacro, SQWangSM, ChengSM} have called into question the basic premises of the tube model, including the picture of barrier-free Rouse retraction. Given the critical role that chain retraction plays in the tube model, a clarification of the molecular relaxation mechanism of entangled polymers after a large step deformation is an urgent need.

Here we present a general approach for extracting microscopic information about molecular relaxation in deformed polymers using small-angle scattering (SAS). By combining and modestly extending the ideas of spherical harmonic decomposition in the literature, we demonstrate how the fingerprint features of molecular relaxation can be obtained by a generalized Fourier analysis of the 2D SAS spectrum. The application of this novel method to small-angle neutron scattering experiments on deformed entangled polymers permits, for the first time, quantitative and model-independent analysis of the full anisotropic 2D spectrum, and provides decisive and convincing evidence against the chain retraction mechanism conceived by the tube model. We show that the two prominent spectral features associated with the chain retraction – peak shift of the leading anisotropic spherical harmonic expansion coefficient and anisotropy inversion in the intermediate wavenumber (\textit{Q}) range around Rouse time ---  were not experimentally observed in a well-entangled polystyrene melt after a large uniaxial step deformation. This result calls for a fundamental revision of the current theoretical picture for nonlinear rheological behavior of entangled polymeric liquids. The application of the spherical harmonic expansion approach, as powerfully illustrated by the current study of entangled polymers, opens a new venue for improving our understanding of macromolecular flow and deformation via rheo-SAS techniques.

\section{Historical Survey of the Field}

The central problem in the study of macromolecular deformation is to gain knowledge about the evolution of conformational statistics of polymers under external perturbation. It has long been recognized that small-angle scattering techniques, particularly small-angle neutron scattering, provide a powerful experimental method for this problem, because of their ability to retrieve microscopic information about chain statistics over a wide range of length scales. The theoretical \cite{Peterlin1957, Peterlin1958, Peterlin1963, Peterlin1964, Peterlin1966, deGennes1974, DE2, SekiyaDoi, Onuki1985, Rabin1988, SQWang1990, Pierleoni1993, Pierleoni1995} and experimental attention \cite{Maconnachie, Boue1, Boue2, Mortensen1, Mortensen2, Schroeder, Blanchard, Muller, Hassager, Linder1985, Linder1988, Linder1989, Hammouda1986, Muller2, Springer1993, Muller1997, Boue2010} in the past, however, has been focused primarily on the analysis of the radius of gyration tensor of a polymer under flow and deformation, and a systematic approach for quantitative analysis of the anisotropic scattering patterns have not emerged from the previous investigations. The radius of gyration, being an averaged statistical quantity, only offers a coarse-grained picture of the molecular deformation on large length scales. In the case of entangled polymers, because of the large overall chain dimensions involved, it often becomes impractical to determine the radius of gyration $R_g$ in a model-independent manner via the Guinier analysis \cite{Higgins}. This difficulty has plagued research aimed at resolving the controversy regarding the chain retraction mechanism proposed by Doi and Edwards. Moreover, the traditional $R_g$ analysis only provides an incomplete picture of the molecular deformation by examining a limited number of directions in space. This method is evidently inadequate in the case of complex scattering patterns \cite{BoueLindner, Straube1995, Read1, Read2, McLeish1999, Pyckhout-Hintzen2001, Pyckhout-Hintzen2004, Read3, Ruocco2013, Ruocco2016, Kirkensgaard}, such as ``butterfly" and ``lozenge" shapes, where a full two-dimensional data analysis is clearly a more desired approach.

Motivated by the aforementioned scientific as well as technical challenges, we set out to explore a different approach to the rheo-SAS problem of polymers, by borrowing, combining, and extending the idea of spherical harmonic expansion (SHE) that has been introduced by several groups of authors in different contexts spanning over a period of roughly half a century \cite{Roe1964, Stuhrmann1970, McBrierty1, McBrierty2, EvansHanley1979, ClarkAckerson, Mitchell1981, Bower, Evans1981, HanleyEvans1982, HessHanley1982, AckersonClark1983, EvansHanleyHess, Mitchell1984, Mitchell1985, Rainwater, Suzuki1987, MWEvans1989, MWEvans, MWEvans1990, Wagner1990, WagnerAckerson, VanGurp, Wagner2002, JFMorris2002, Wagner2015}.

Building on the Taylor expansion treatment of earlier researchers \cite{IrvingKirkwood1950, HSGreen, SniderCurtiss}, Evans, Hanley, and Hess \cite{EvansHanley1979, Evans1981, HanleyEvans1982, HessHanley1982, EvansHanleyHess, Rainwater} were among the first who systematically investigated the structural distortion of simple fluids under shear by expressing the anisotropic pair distribution function in terms of spherical harmonics. These computational studies inspired the discussion of the principles of group-theoretical statistical mechanics for non-Newtonian flow \cite{MWEvans1989, MWEvans, MWEvans1990}, and these concepts were also echoed by the experimental efforts of a number of research groups \cite{ClarkAckerson, AckersonClark1983, Suzuki1987, Wagner1990, WagnerAckerson} around the same time. However, these investigations focused exclusively on colloidal suspensions under small shear perturbation, whereas large extensional deformation is the preferred condition for probing polymeric systems. Additionally, while it is straightforward to perform spherical harmonic decomposition in computer simulations where three-dimensional real space information of particle coordinates is readily available, small-angle scattering experiments can only access the two-dimensional reciprocal space cross sections on certain planes, for which the projected spherical harmonics may not necessarily form an orthogonal basis set. The delicacy of this issue has not been fully appreciated until very recently \cite{WRChen}.

In the polymer community, Roe and Krigbaum have already conceived the idea of spherical harmonic expansion of the orientation distribution function of statistical segments in deformed polymer networks and discussed the potential application of this technique in analyzing the variation of x-ray intensity of the “amorphous halo” observed for stretched polymers \cite{Roe1964}. However, it was not until the work of Mitchell and coworker almost twenty years later \cite{Mitchell1981, Mitchell1984, Mitchell1985}, that a more formal treatment of the measured scattering intensity in terms of Legendre expansion for the uniaxial extensional geometry was developed. Despite the widespread use of this method, the polymer community has so far mainly looked at the problem of scattering of deformed polymers through the lens of rheology, where the major interest is to extract an order parameter to compare with stress. Consequently, the previous works in this area fell short at recognizing the value of spherical harmonic expansion as a general approach for characterizing \textit{Q}-dependent deformation anisotropy and chain conformation at different length scales.

\section{Spherical Harmonic Expansion Approach}
\subsection{A Philosophical Shift}

In this section, we present our general formulation of the small-angle scattering problem of deformed polymers. We will start the discussion by describing the angle from which we approach this topic. As we shall demonstrate, our viewpoint represents a \textit{philosophical} departure from the previous method employed in the polymer community, where the primary concern was to extract a single order parameter.
Following the convention in the field of polymer dynamics \cite{DE, Bird2}, let us suppose the polymer chain is modelled by a series of $N$ beads, each located at $\bm{r}_i$. In the context of small-angle neutron scattering by isotopically labelled deformed melts, the measured coherent scattering intensity $I_{\mathrm{coh}}$, which is dependent on scattering wave vector $\bm{Q}$, is proportional to the \textit{single-chain} structure factor (form factor) $S(\bm{Q})$ \cite{Higgins, deGennesBook}:
\begin{eqnarray}
\begin{aligned}
I(\bm{Q}) &=I_{\mathrm{coh}}(\bm{Q})+I_{\mathrm{inc}}\\
&=(b_D-b_H)^2 f(1-f)nN^2S(\bm{Q})+I_{\mathrm{inc}},
\end{aligned}
\end{eqnarray}
\begin{equation}
S(\bm{Q})=\frac{1}{N^2}\langle \sum_{i,j}^N e^{-i \bm{Q}\cdot (\bm{r}_i -\bm{r}_j)} \rangle,
\end{equation}
where ($b_D-b_H$) is the contrast factor due to the difference in the coherent scattering length between hydrogen and deuterium, $f$ the fraction of the labeled species, n the number density, and $I_{\mathrm{coh}}$ the incoherent background. $\langle \ldots \rangle$ stands for ensemble average. Let $\psi_{ij}(\bm{r})$ be the segment distribution function that describes statistically the separation between beads $i$ and $j$. We could define an intra-chain pair distribution function $g(\bm{r})$ as \cite{Zimm}:
\begin{equation}
g(\bm{r})=\frac{1}{N^2}\sum_{i=1}^N \sum_{j=1,j\neq i}^N \psi_{ij}(\bm{r}),
\end{equation}
which is related to the single-chain structure factor through the Fourier transform:
\begin{equation}
S(\bm{Q})=\int g(\bm{r}) e^{-i \bm{Q}\cdot \bm{r}} d\bm{r}.
\end{equation}

We would like to note that statistical distribution functions are the centerpiece of the kinetic theory of polymer fluid dynamics. The long tradition of kinetic theory for polymeric liquids, was initiated by the celebrated paper of Kramers \cite{Kramers}, developed by Kirkwood \cite{KirkwoodRiseman, IrvingKirkwood1950, Kirkwood1954}, Rouse \cite{Rouse}, Zimm \cite{Zimm2}, Lodge \cite{ASLodge, ASLodge2, LodgeWu}, Yamamoto \cite{Yamamoto1}, and others \cite{Hassager1974, KingJames, CurtissBird1, CurtissBird2}, and epitomized in the classical book by Bird \textit{et al}. \cite{Bird2}. We see, from Eqs. 1-4, that the spatially anisotropic scattering intensity accessed by SAS techniques in the reciprocal space reflects nothing but the perturbation of configuration distribution functions of the polymer chain by external deformation. However, this seemingly obvious and yet fundamental viewpoint has not been fully appreciated, as witnessed by the immense disparity between theoretical development and experimental efforts by SAS. As we shall show below, a powerful weapon for analyzing SAS data can be forged by drawing upon the concept of spherical harmonic expansion. This new approach supplies a convenient platform for connecting small-angle scattering experiments and statistical and molecular theories of polymers.

\subsection{3D Decomposition and 2D Reconstruction}

In the context of our current investigation, the measured scattering signal is dominated by coherent scattering, \textit{i.e.}, $I_{\mathrm{coh}}\gg I_{\mathrm{inc}}$. Thus,
\begin{equation}
S(\bm{Q})=I(\bm{Q})/\left[ \lim_{Q \to 0}I_{\mathrm{iso}}(Q) \right],
\end{equation}
where $I_{\mathrm{iso}}(Q)$ is the scattering intensity from the isotropic sample. Because of this simple proportionality between $S(\bm{Q})$ and $I(\bm{Q})$, we will focus only on $S(\bm{Q})$ in the discussions below.

\begin{figure*}
\centering
\includegraphics[scale=0.5]{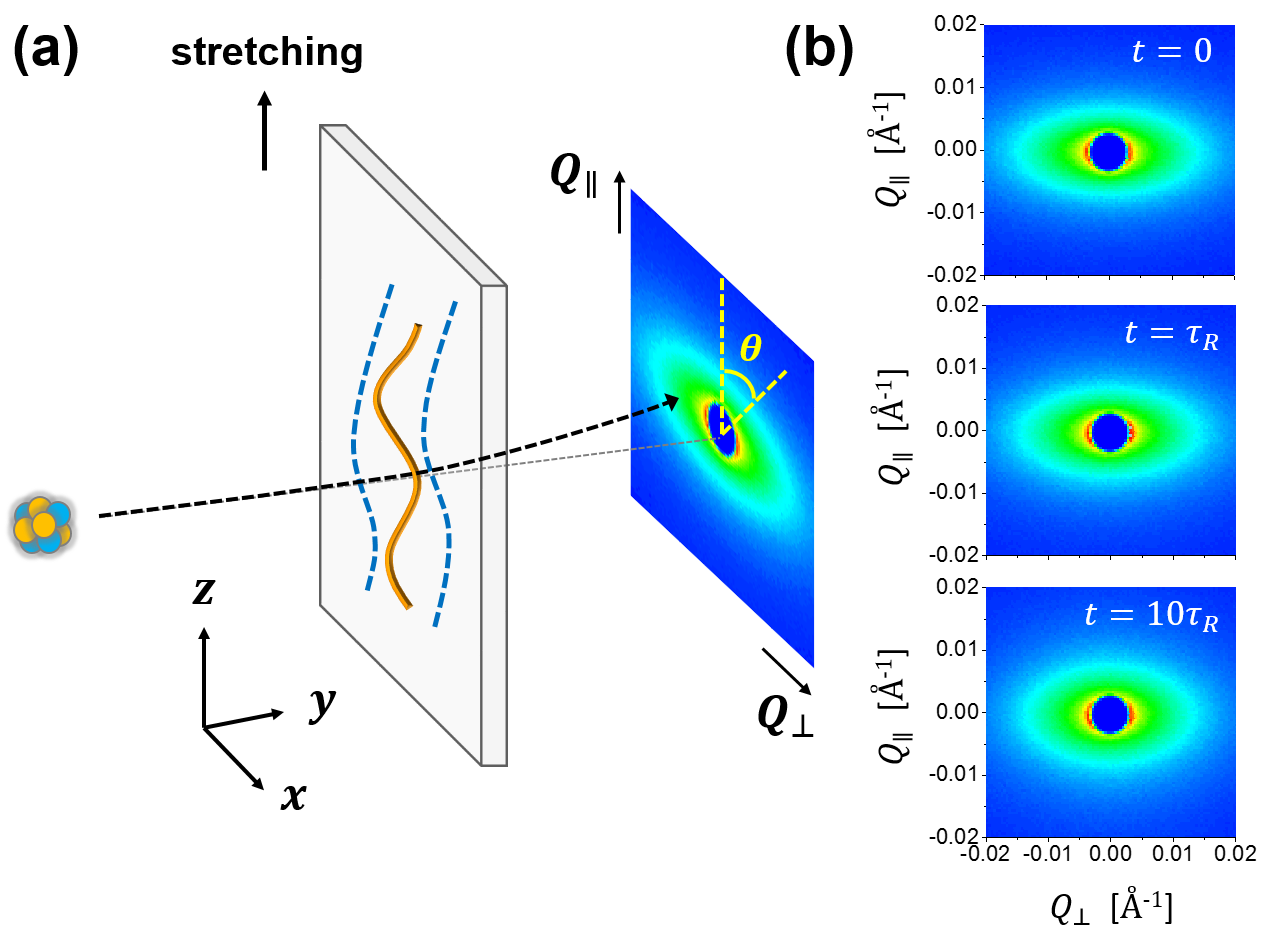}
\caption{(a) Illustration of SANS measurements on uniaxially elongated samples: the stretching is along the \textit{z}-axis whereas the incident SANS beam is perpendicular to the \textit{xz}-plane. (b) Evolution of the SANS spectrum with polymer relaxation.}
\label{F1}
\end{figure*}

Formally, the dependence of the single-chain structure factor (form factor) $S(\bm{Q})$ on the magnitude ($Q$) and orientation ($\bm{\Omega}$) of the scattering wave vector can be expressed in terms of spherical harmonics:
\begin{equation}
S(\bm{Q})=\sum_{l,m}S_{l}^{m}(Q)Y_{l}^{m}(\bm{\Omega}),
\end{equation}
where $S_l^m(Q)$ is the expansion coefficient corresponding to each real spherical harmonic function $Y_l^m(\bm{\Omega})$. In this work, our choice of the spherical coordinates follows the convention in physics \cite{Cahill}, for which $\theta$ is the polar angle from the positive \textit{z}-axis with $\theta \in [0,\pi]$, and $\phi$ is the azimuthal angle in the \textit{xy}-plane from the \textit{x}-axis  with $\phi \in [0,2\pi)$. For the uniaxial extension problem investigated herein, the stretching is along the \textit{z}-axis and the incident neutron beam is perpendicular to the \textit{xz}-plane (Fig. \ref{F1}a). Our real spherical harmonic functions are defined as:
\begin{widetext}
\begin{equation}
Y_l^{m}(\bm{\Omega})=Y_l^m(\theta,\phi)=\begin{cases}
\sqrt{2}\sqrt{(2l+1)\frac{(l-|m|)!}{(l+|m|)!}}P_l^{|m|}(\cos\theta)\sin(|m|\phi)      & \quad (m<0)\\
\sqrt{2l+1}P_l^0(\cos\theta)  & \quad (m=0)\\
\sqrt{2}\sqrt{(2l+1)\frac{(l-m)!}{(l+m)!}}P_l^{m}(\cos\theta)\cos(m\phi)       & \quad (m>0)
\end{cases}
\end{equation}
\end{widetext}
which differ from the classical definitions by a factor of $1/\sqrt{4\pi}$. Because of the axial symmetry of the uniaxial extension problem, it is easy to see that all the $m\neq 0$ terms and the odd $l$ terms are forbidden \cite{Roe1964, Mitchell1981, Rainwater, JFMorris2002}, namely:
\begin{eqnarray}
\begin{aligned}
S(\bm{Q}) &=S(Q,\theta)\\
&=\sum_{l\mathrm{:even}}S_l^0(Q)Y_l^0(\theta)\\
&=\sum_{l\mathrm{:even}}S_l^0(Q)\sqrt{2l+1}P_l^0(\cos\theta).
\end{aligned}
\end{eqnarray}
In other words, $S(\bm{Q})$ is independent of $\phi$ and could be written as a linear combination of even order Legendre functions. In particular, the term $S_0^0 (Q)Y_0^0(\theta)$ represents the isotropic part of the distorted structure, whereas the term $S_2^0(Q)Y_2^0(\theta)$ is the leading anisotropic component that corresponds to the symmetry of uniaxial deformation.

Equation 8 gives the spherical harmonic expansion of the anisotropic single-chain structure factor in \textit{three-dimensional} space. In order to obtain the expansion coefficients $S_l^0(Q)$ from SAS experiments, we must consider the cross section of $S(\bm{Q})$ on the \textit{two-dimensional} detector plane. In the case of shear, the low symmetry of this geometry makes the reconstruction of $S(\bm{Q})$ from 2D scattering patterns rather complicated \cite{WRChen}. However, the unique symmetry of uniaxial extension greatly simplifies the matter. It is evident from Eq. 8 that, the cross section of $S(\bm{Q})$ on the \textit{xz}-plane is:
\begin{eqnarray}
\begin{aligned}
S(Q_x,Q_y=0,Q_z)&=S(Q,\theta,\phi=0)=S(Q,\theta)\\
&=\sum_{l\mathrm{:even}}S_l^0(Q)Y_l^0(\theta).
\end{aligned}
\end{eqnarray}
Equation 9 and the fact that $\int_{-1}^{1}P_n^0(x)P_m^0(x)dx=\frac{2}{2n+1}\delta_{nm}$ indicate that, $Y_l^0(\theta)$ form an orthogonal basis set not only in 3D space but also on the \textit{xz}-plane. Therefore, the expansion coefficient $S_l^0(Q)$ can be straightforwardly computed from the small-angle scattering pattern on the \textit{xz}-plane as:
\begin{eqnarray}
\begin{aligned}
S_l^0(Q)&=\frac{1}{2}\int_0^{\pi}S(Q,\theta,\phi=0)Y_l^0(\theta)\sin \theta d\theta\\
&=\frac{1}{2\lim_{Q\to 0}I_{\mathrm{iso}}(Q)}\int_{0}^{\pi}I_{xz}(Q,\theta)Y_l^0(\theta)\sin \theta d\theta,
\end{aligned}
\end{eqnarray}
where $I_{xz}(Q,\theta)$ is the detected scattering intensity on the \textit{xz}-plane. 

In passing, we note that the spherical harmonic expansion approach is inclusive of the traditional data analysis method that focuses on the scattering intensities along the parallel and perpendicular directions: the projected structures in these directions could be expressed as linear combinations of the expansion coefficients. For example, we have:
\begin{widetext}
\begin{eqnarray}
\begin{aligned}
S_{\parallel}(Q)&=S(Q_x=0,Q_y=0,Q_z)=S(Q,\theta=0)=\sum_{l:\mathrm{even}}S_l^0(Q)\sqrt{2l+1}P_l^0(1)\\
&=S_0^0(Q)+\sqrt{5}S_2^0(Q)+\sqrt{9}S_4^0(Q)+\sqrt{13}S_6^0(Q)+\sqrt{17}S_8^0(Q)+\ldots
\end{aligned}
\end{eqnarray}
\begin{eqnarray}
\begin{aligned}
S_{\perp}(Q)&=S(Q_x,Q_y=0,Q_z=0)=S(Q,\theta=\frac{\pi}{2})=\sum_{l:\mathrm{even}}S_l^0(Q)\sqrt{2l+1}P_l^0(0)\\
&=S_0^0(Q)-\frac{\sqrt{5}}{2}S_2^0(Q)+\frac{9}{8}S_4^0(Q)-\frac{5\sqrt{13}}{16}S_6^0(Q)+\frac{35\sqrt{17}}{128}S_8^0(Q)+\ldots
\end{aligned}
\end{eqnarray}
\end{widetext}
where $S_{\parallel}(Q)$ and $S_{\perp}(Q)$ are the cross sections of $S(\bm{Q})$ along the parallel and perpendicular directions to stretching, respectively.

\subsection{Fingerprinting Molecular Deformation}

\begin{figure*}
\centering
\includegraphics[scale=0.48]{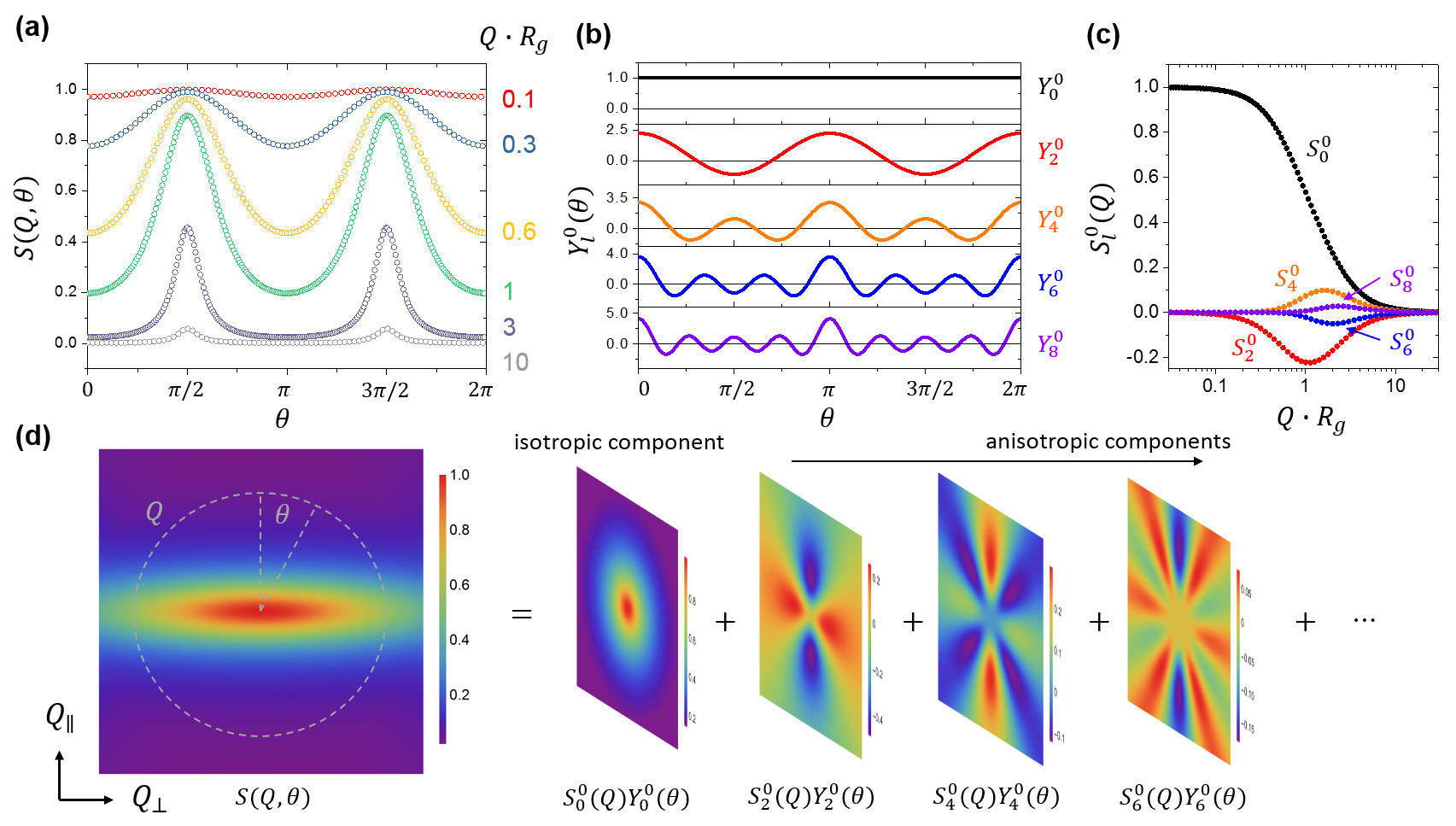}
\caption{Illustration of the spherical harmonic expansion approach with the simulated spectrum from the affine model, for a linear polymer that is uniaxially stretched to $\lambda=3$. (a) Angular dependence of the anisotropic single-chain structure factor at various \textit{Q}s. (b) Angular dependence of the projections of the spherical harmonic functions on the \textit{xz}-plane. As discussed in the text, these are essentially Legendre functions. (c) The \textit{Q}-dependent expansion coefficients $S_l^0(Q)$ are given by Legendre expansion of $S(Q,\theta)$. (d) The spherical harmonic expansion decomposes the 2D SANS spectrum into contributions from different symmetries: the isotropic component $S_0^0(Q)Y_0^0(\theta)$, and the anisotropic components $S_2^0(Q)Y_2^0(\theta)$, $S_4^0(Q)Y_4^0(\theta)$, $S_6^0(Q)Y_6^0(\theta)$, \textit{etc}.}
\label{F2}
\end{figure*}

To further illustrate the idea of spherical harmonic expansion analysis, let us consider a simulated single-chain structure factor for a polymer uniaxially elongated to a stretching ratio $\lambda$ of 3.0 (Fig. \ref{F2}), calculated using the affine model \cite{Ullman1979, Hassager}. At a given magnitude of the scattering wave vector, $Q$, $S(\theta)$ is a periodic function of $\theta$ with a period of $\pi$ (Fig. \ref{F2}a). Because of the orthogonality of $Y_l^0(\theta)$, $S(\theta)$ can be decomposed in terms of $Y_l^0(\theta)$, and the expansion coefficient $S_l^0$ can be obtained by angular averaging $S(\theta)$ with the weighing factor $Y_l^0(\theta)$ (Eq. 10, Figs \ref{F2}b and \ref{F2}c). Carrying out this procedure for all the different \textit{Q}s, we translate the anisotropic 2D scattering pattern (Fig. \ref{F2}d) into a 1D plot of \textit{Q}-dependent expansion coefficients $S_l^0(Q)$ (Fig. \ref{F2}c).

While Fig. \ref{F2}c and Fig. \ref{F2}d contain the same amount of information mathematically, the plot of expansion coefficients is much more convenient to analyze in great details. Moreover, by isolating the spectral contributions from different symmetries (Fig. \ref{F2}d), the spherical harmonic decomposition approach provides a new means to study the molecular relaxation/deformation mechanisms of polymers, as we shall see in Sections IV and V. 

Mathematically, our treatment of the small-angle scattering spectrum can be regarded as a generalized Fourier expansion approach. The anisotropic single-chain structure factor is decomposed by spherical harmonic functions and re-synthesized from the 2D patterns in small-angle scattering experiments. This approach helps to distill the ``hidden" information about molecular deformation from the distorted 2D spectrum. At this point, it is useful to draw an analogy to the ideas of ``rheological fingerprinting" complex fluids using large-amplitude oscillatory shear \cite{DodgeKrieger, Giacomin, Wilhelm1998, Wilhelm1999a, Kallus, Wilhelm2002, Hyun2007, Ewoldt2007a, Ewoldt2007b, Ewoldt2008, Hyun2009, Ewoldt2010, Hyun2011, Rogers2012, McKenna2014}. In particular, it has been proposed that the Fourier or Chebyshev expansion coefficients for the stress response could be used to define unique fingerprints of nonlinear rheology of soft viscoelastic materials and reveal properties that are typically obscured by conventional test protocols. It has also been recognized that the model-independent nature of the harmonic analysis not only enables quantitative characterization of materials, but also allows one to challenge constitutive relations. From this perspective, our spherical harmonic expansion approach to SAS and the widely used (generalized) Fourier analysis in the complex fluids community share a similar philosophical root.

\section{Molecular Fingerprints of Chain Retraction}

\begin{figure*}
\centering
\includegraphics[scale=0.6]{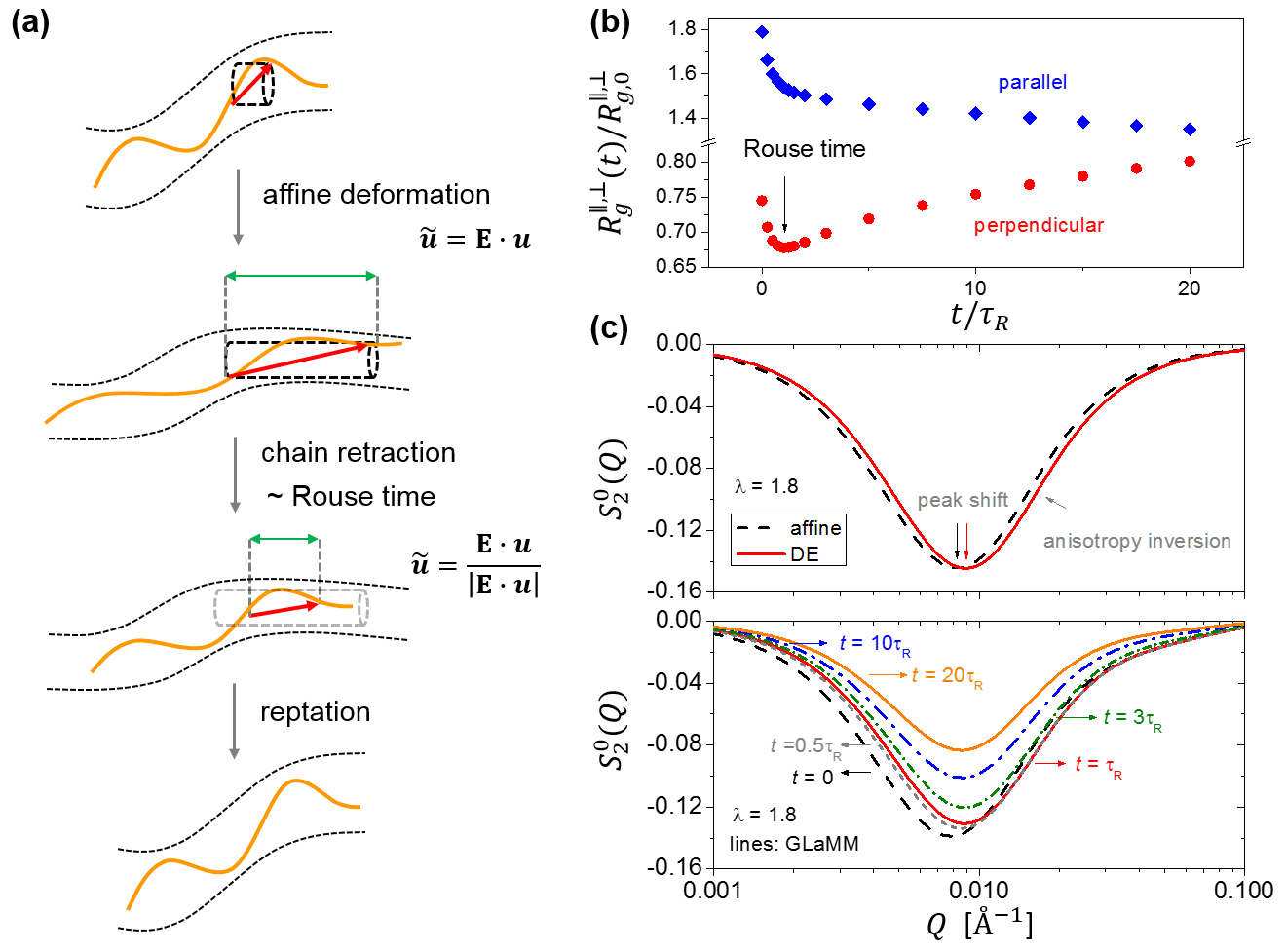}
\caption{(a) Illustration of the molecular relaxation mechanism envisioned by the Doi-Edwards theory. The chain conformation immediately after the step-strain deformation can be described by the affine transformation. The chain retraction around Rouse time quickly equilibrates the contour length and leads to a reduction of all components of the radius of gyration tensor. The molecular relaxation continues via reptation after chain retraction. (b) Evolution of the radius of gyration in the parallel and perpendicular directions to stretching, as predicted by the GLaMM model for an entangle polystyrene ($Z=34$) after a step deformation of $\lambda=1.8$, performed with a constant crosshead velocity $v=40 l_0/\tau_R$. (c) Upper panel: Expansion coefficient $S_2^0(Q)$ before and after the chain retraction, calculated using the affine model and Doi-Edwards (DE) model, respectively. Lower panel: Predictions from the GLaMM model. Our choice of the GLaMM parameters follows the standard practice in the literature \cite{Graham}.}
\label{F3}
\end{figure*}

Having laid down the foundation for the spherical harmonic expansion technique, let us now return to the central question that we raised at the beginning of this article: how can we critically test the chain retraction hypothesis of the tube theory for entangled polymers? 
The investigations in the past have been focused on the analysis of the radius gyration tensor in step-strain relaxation experiments, following the original strategy outlined in the celebrated 1978 paper of Doi and Edwards \cite{DE2} (Fig. \ref{F3}a). Theoretically, immediately after a fast step deformation, the radius of gyration tensor $\langle R_{g}^2 \rangle_{\alpha\beta}$ is equal to the affinely deformed one \cite{DE2, SekiyaDoi}:
\begin{equation}
\langle R_{g}^2 \rangle_{\alpha\beta}=\langle R_{g}^2 \rangle_0 \langle \left(  \mathbf{E}\cdot \bm{u}\right)_{\alpha} \cdot \left(  \mathbf{E}\cdot \bm{u}\right)_{\beta} \rangle
\end{equation}
where $\langle R_{g}^2 \rangle_0$ is the equilibrium mean-square radius of gyration, $\mathbf{E}$ is the deformation gradient tensor, and $\bm{u}$ is a unit vector of isotropic distribution. The averaging $\langle \ldots \rangle$ for $\left(  \mathbf{E}\cdot \bm{u}\right)_{\alpha} \cdot \left(  \mathbf{E}\cdot \bm{u}\right)_{\beta}$ is taken over the equilibrium distribution of $\bm{u}$. The chain retraction along the tube around the Rouse time would reduce all components of $\langle R_{g}^2 \rangle_{\alpha\beta}$ by a factor of $\langle |\mathbf{E}\cdot \bm{u}| \rangle$:
\begin{equation}
\langle R_{g}^2 \rangle_{\alpha\beta}=\langle R_{g}^2 \rangle_0 \frac{\langle \left(  \mathbf{E}\cdot \bm{u}\right)_{\alpha} \cdot \left(  \mathbf{E}\cdot \bm{u}\right)_{\beta} \rangle}{\langle |\mathbf{E}\cdot \bm{u}| \rangle}
\end{equation}
After the retraction, the chain continues to relax towards the equilibrium state through reptation. In the case of uniaxial extension geometry, the above-mentioned mechanism is expected to lead to a non-monotonic change of radius of gyration in the perpendicular direction during the stress relaxation.

Figure \ref{F3}b gives an example for the evolution of the radius of gyration in the parallel and perpendicular directions to stretching, calculated according to the modified tube model proposed by Graham, Likhtman, Milner, and McLeish \cite{Graham}, \textit{i.e.}, the GLaMM model. The GLaMM model is widely considered the-state-of-the-art version of the tube theory, as it incorporates the effects of reptation, chain stretch, and convective constraint release on the microscopic level through a stochastic partial differential equation for the contour dynamics. From Fig. \ref{F3}b, we see that the qualitative feature of chain retraction --- the non-monotonic change of $R_g$ in the perpendicular direction --- is well captured by the GLaMM model. In addition, the magnitude of retraction, \textit{i.e.}, $R_{g}^{\perp}(t=0)/R_{g}^{\perp}$, is also consistent with the expectation from the original Doi-Edwards theory, which predicts the ratio to be $\sqrt{\langle |\mathbf{E}\cdot \bm{u}| \rangle}$.

In principle, one should be able to critically test the chain retraction hypothesis by performing SANS experiments on uniaxially-stretched entangled polymer melts and comparing the measured $R_g$ with theoretical predictions. In reality, experimentalists have encountered tremendous difficulty in following this approach. First, due to inevitable elastic breakup after a large step deformation \cite{YYW2007}, stress relaxation experiments of this kind are typically restricted to relatively small strains. This constraint means the magnitude of chain retraction would be rather small and thus require highly accurate SANS experiments. On the other hand, however, it is practically impossible to reliably determine the radius of gyration tensor through \textit{model-independent} Guinier analysis, because of the limited \textit{Q} range and flux of existing SANS instruments and the large molecular size of entangled polymers. As a result, experimentalists in the past have had to resort to using the affine-like model to determine $R_g$ by averaging over an opening angle along the principal axes \cite{Boue1, Hammouda1986, Mortensen1, Muller, Schroeder, Blanchard}. Putting the ambiguity in model fitting aside, this approach does not even seem to be logically self-consistent: it is not possible to critically test a non-affine model (tube model) by fitting the experimental data with the affine-like model which assumes the same transformation rule for chain conformation at all length scales. 

To circumvent the dilemma with the traditional $R_g$ analysis, here we propose a different approach to examine the chain retraction hypothesis by using the spherical harmonic expansion technique. The upper panel of Fig. \ref{F3}c presents the major component of the deformation anisotropy $S_2^0(Q)$ before and after the chain retraction for a step strain of $\lambda=1.8$, calculated using the affine model and Doi-Edwards model \cite{DE2}, respectively. We see that the chain retraction would lead to a horizontal shift of $S_2^0(Q)$ towards higher $Q$. This prediction is consistent with the physical picture offered by the tube model: the chain retraction reduces the overall dimension of the chain, causing the horizontal shift, but the orientation anisotropy is not relaxed, as the peak amplitude remains the same. This analysis shows that there are two distinct spectral features associated with the chain retraction in a step-strain relaxation experiment: the peak shift of $S_2^0(Q)$ and the increase of anisotropy in the intermediate $Q$ range. We term the latter feature ``anisotropy inversion" --- instead of relaxation of deformation anisotropy, the chain retraction is expected to give rise to an increase of anisotropy in the intermediate $Q$ range.

Having made this qualitative analysis with the original Doi-Edwards theory, we now turn to the more sophisticated GLaMM model for quantitative predictions (Fig. \ref{F3}c, lower panel). First, the GLaMM model still faithfully captures the two unique features of chain retraction, \textit{i.e.}, peak shift and anisotropy inversion. Moreover, the original Doi-Edwards model and the GLaMM model produce consistent calculations about the magnitude of the peak shift. Beyond Rouse time, the GLaMM model predicts that $S_2^0(Q)$ continues to relax towards the equilibrium state without much change in the peak position, in agreement with the idea that relaxation after chain retraction is orientational. 

The above calculations and analyses powerfully demonstrate that the spherical harmonic expansion technique allows us to directly translate the physical idea of chain retraction into unique and intuitive spectral patterns. More importantly, it provides a platform for us to bring together theory and experiment, and to critically test, for the first time, the retraction hypothesis in a model-independent, nonlinear-fitting-free manner.

\section{New Results and Discussions}
\subsection{Experimental Methods}

Equipped with the new insight from spherical harmonic expansion analysis, we carried out a critical examination of the chain retraction hypothesis of the tube model, by using small-angle neutron scattering. Our experimental system was based on the mixture of protonated and deuterated polystyrene (PS) homopolymers that were synthesized by anionic polymerization in benzene with \textit{sec}-butyllithium as the initiator (\textit{h}-PS: $M_w = 450$kg/mol, $M_w/M_n = 1.06$; \textit{d}-PS: $M_w = 510$ kg/mol, $M_w/M_n = 1.04$). The \textit{h}-PS and \textit{d}-PS were dissolved at an \textit{h}/\textit{d} ratio of 5/95 in toluene, fully mixed, and precipitated in excess methanol. The resulting blend was dried in a vacuum oven first at room temperature and then at $130^{\circ}$C to completely remove the residual solvents.

\begin{figure*}
\centering
\includegraphics[scale=0.52]{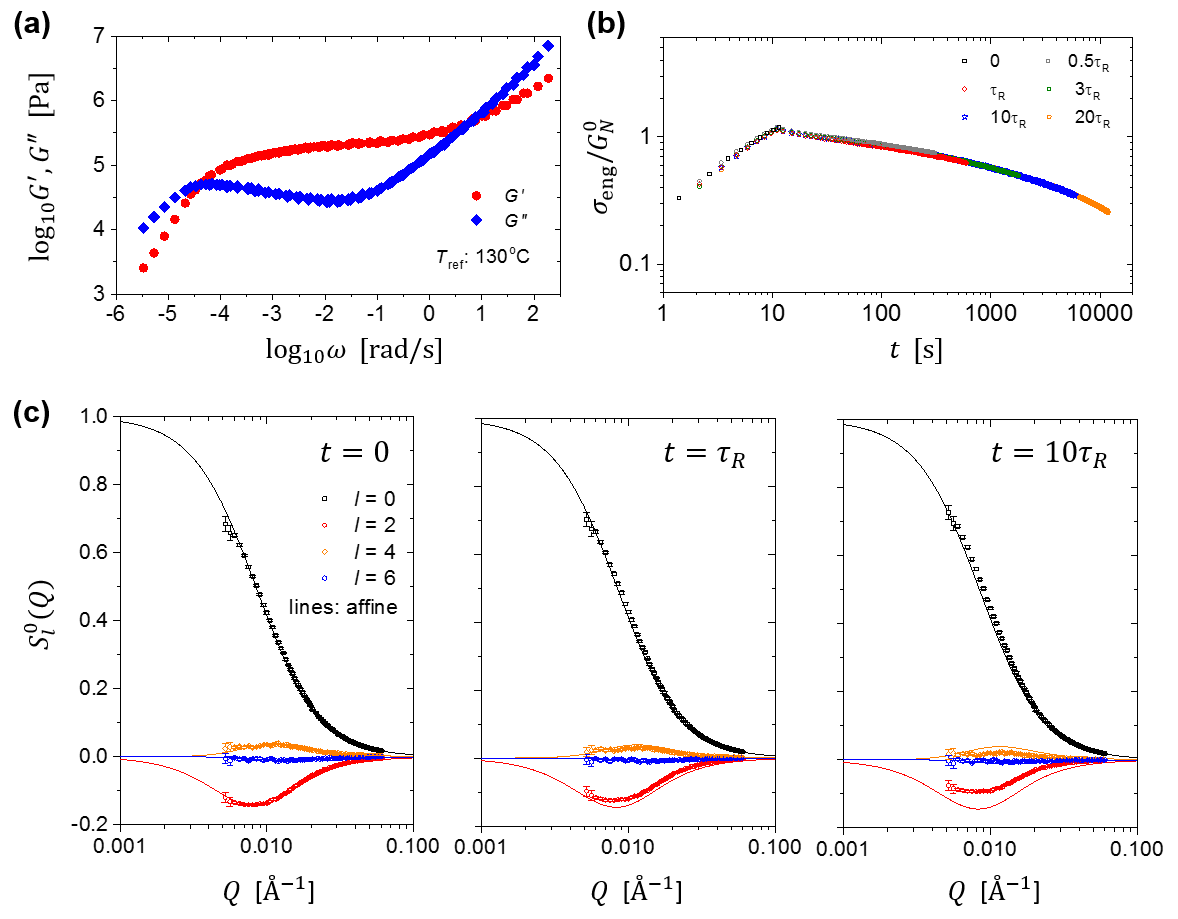}
\caption{(a) Linear viscoelastic properties of the mixture of \textit{d}-PS and \textit{h}-PS at $130^{\circ}$C. (b) Stress relaxation behavior after a step deformation of $\lambda=1.8$, performed with a constant crosshead velocity $v=40 l_0/\tau_R$. (c) Expansion coefficients $S_l^0(Q)$ at $t=0$, $\tau_R$, and $10\tau_R$. The solid lines are computed according to the affine deformation model for $\lambda=1.8$. The SANS data was collected on the NGB30 SANS beamline at NIST.}
\label{F4}
\end{figure*}

The linear viscoelastic properties of the blend were characterized on an HR2 rheometer (TA Instruments) by small amplitude oscillatory shear measurements in the frequency range 0.1-–100 rad/s and at temperatures between 200 and $120^{\circ}$C. Figure \ref{F4}a shows the master curve for the dynamic moduli ($G^{\prime}$ and $G^{\prime\prime}$) at $130^{\circ}$C, constructed by using the time-temperature superposition principle \cite{Ferry}. The average number of entanglements per chain, $Z$, is estimated to be 34 for this system ($Z=G_e M_w/\rho RT$, with $G_e$ being the plateau modulus and $\rho$ the polymer density). We evaluated the Rouse relaxation time, $\tau_R$, using three different methods, the classical tube model formula ($\tau_R=\tau/3Z$, with $\tau$ being the reptation time) \cite{RubinsteinColby}, the Osaki formula [$\tau_R=(6M_w \eta/\pi^2 \rho RT)\cdot (1.5M_e/M_w)^{2.4}$, with $\eta$ being the zero-shear viscosity and $M_e$ the entanglement molecular weight] \cite{Osaki1982, Osaki2001}, and the Likhtman-McLeish theory \cite{LMc}, which yield 251 s, 592 s, and 715 s, respectively at $130^{\circ}$C. In this work, we choose to use Osaki's formula as it overcomes the well-known problem with the classical tube model formula and is yet much more straightforward than the Likhtman-McLeish theory.

The specimens for the SANS measurements were prepared on an RSA-G2 Solids Analyzer from TA Instruments (Fig. \ref{F4}b). The temperature was controlled by the Forced Convection Oven of the RSA-G2, using nitrogen as the gas source. Rectangular samples were uniaxially stretched at $130^{\circ}$C to $\lambda=1.8$ with a constant crosshead velocity $v=40 l_0/\tau_R$, where $l_0$ is the initial length of the sample. The samples were allowed to relax for different amount of time (from 0 to $20\tau_R$) at $130^{\circ}$C and then immediately quenched by pumping cold air into the oven. At $130^{\circ}$C, the Rouse time of the sample was about 10 minutes, whereas the terminal relaxation time is on the order of 7 hours. Furthermore, since the test temperature is only about $30^{\circ}$C above $T_g$, the relaxation time increases sharply with decreasing temperature. In our experiment, it took less than 10 s for the temperature to drop from $130^{\circ}$C to $125^{\circ}$C at which point the chain relaxation was already exceedingly slow. Therefore, we were able to effectively freeze the conformation of the polymer chain with negligible stress relaxation during the quenching procedure.

Small-angle neutron scattering measurements of the quenched glassy polystyrene films were performed on the NGB30 SANS diffractometers at the Center for Neutron Research of NIST. Two wavelengths of incident neutrons, 6.0 and 8.4 \r{A}, were used to cover a range of scattering wave vector $Q$ from 0.001 to 0.1 \r{A}$^{-1}$. The measured intensity was corrected for detector background and sensitivity, and placed on an absolute scale using a direct beam measurement.

\subsection{Spherical Harmonic Expansion Analysis}

Figure \ref{F4}c presents spherical harmonic expansion coefficients $S_l^0(Q)$ ($l=0,2,4,6$) immediately after the step deformation ($t=0$), and at $\tau_R$ and $10\tau_R$. As a reference, we also plot the coefficients of the affine deformation model for $\lambda=1.8$. First, Fig. \ref{F4}c nicely illustrates the benefit of performing spherical harmonic decomposition. The isotropic component $S_0^0(Q)$, which does not change significantly from $t=0$ to $t=10\tau_R$, makes a major contribution to the 2D spectrum. On the other hand, the relaxation of the anisotropic coefficients $S_2^0(Q)$ and $S_4^0(Q)$ is clearly visible during the same period of time. Therefore, it makes sense to separate these different components via the spherical harmonic decomposition technique, rather than directly perform analysis on the composite 2D spectra (Fig. \ref{F1}b), which do not exhibit any characteristic features. Second, the affine model seems to give a satisfactory description of the molecular deformation during the step uniaxial stretching (Fig. \ref{F4}c, left panel), although upon closer examination, we do find that the affine model slightly overestimates the anisotropy at high \textit{Q}s (not visible on the scale of the current plot). This result should be expected, because the step deformation was performed with a high strain rate –-- the initial Rouse Weissenberg number ($\tau_R v/l_0$) in this case was 40.

\begin{figure*}
\centering
\includegraphics[scale=0.4]{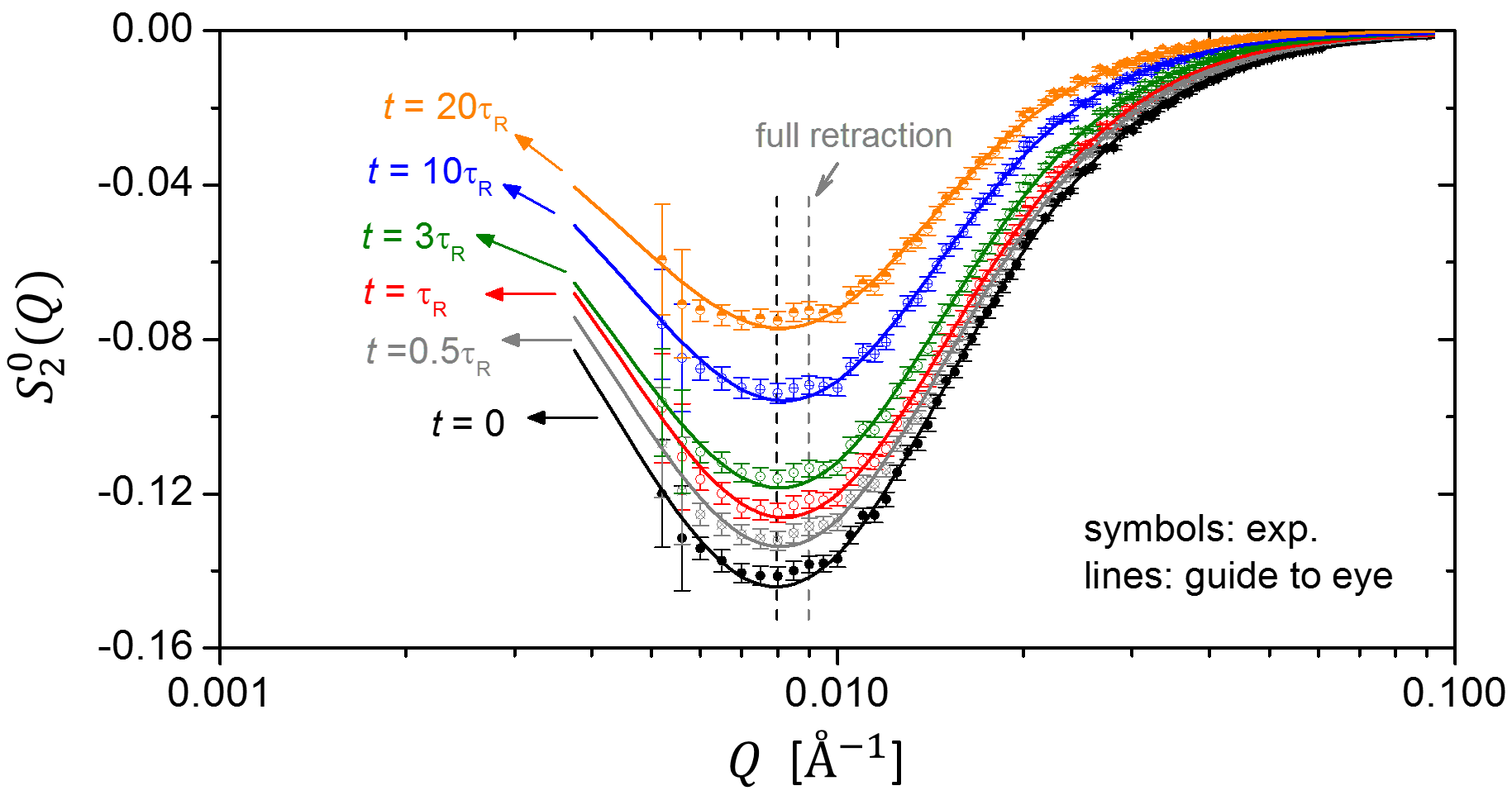}
\caption{Evolution of the expansion coefficient $S_2^0(Q)$. The vertical grey dashed line indicates the theoretically expected peak position after full chain retraction. The SANS data was collected on the NGB30 SANS beamline at NIST.}
\label{F5}
\end{figure*}

Our analysis in Section IV reveals that the chain retraction mechanism should give rise to two distinct spectral features for the leading anisotropic component $S_2^0(Q)$: peak shift and anisotropy inversion. Now let us turn to this critical test of the retraction hypothesis and examine the evolution of $S_2^0 (Q)$ during the stress relaxation. Figure \ref{F5} shows the expansion coefficients $S_2^0(Q)$ at $t=0$, $0.5\tau_R$, $\tau_R$, $3\tau_R$, $10\tau_R$, and $20\tau_R$. The black dashed line marks the peak position immediately after the step deformation ($t=0$), whereas the grey dashed line indicates the theoretically expected peak position after full chain retraction, at $t\approx\tau_R$. As we pointed out in Section IV, the ``simple" Doi-Edwards model and the more sophisticated GLaMM model give the same prediction for the peak position after retraction.

It is evident from Fig. \ref{F5} that the unique scattering patterns (Fig. \ref{F3}c) associated with chain retraction were not experimentally observed. The peak shift is negligible up to 20 times of Rouse relaxation time, suggesting there is no strong decoupling of stretch and orientation relaxation. Furthermore, there is no anisotropy inversion either: the anisotropy decays monotonically with time at all \textit{Q}s. Here we emphasize the model-independent nature of the spherical harmonic expansion analysis --- it is simply a different way of presenting the ``raw" 2D data. Unlike the previous investigations, there is no ambiguity associated with model fitting and no room for human bias. Therefore, our critical test clearly demonstrates that the chain retraction hypothesis of the tube model is not supported by small-angle neutron scattering experiments.

\subsection{Analysis of Radii of Gyration}

Having reviewed the evidence from spherical harmonic expansion analysis, let us now return to the traditional analysis of the radius of gyration tensor. As we have explained in Sections II and IV, it is not feasible to extract $R_g$ from the SANS measurement using model-independent Guinier analysis due to the large size of the polymer chain. Following the common procedure in the literature, we apply a modified Debye function to determine the $R_g$ in the parallel and perpendicular directions to stretching:
\begin{eqnarray}
\begin{aligned}
I=&2I_0\left[e^{-\left(Q_{\parallel,\perp}R_g^{\parallel,\perp}\right)^2}+\left(Q_{\parallel,\perp}R_g^{\parallel,\perp}\right)^2-1\right] \\ 
&/\left(Q_{\parallel,\perp}R_g^{\parallel,\perp}\right)^4+I_{\mathrm{inc}},
\end{aligned}
\end{eqnarray}
where $I_0$ is the forward scattering intensity and $I_{\mathrm{inc}}$ is the incoherent background. However, we would like to stress that Eq. 15 should only be taken as an approximate form for the scattering intensity in the intermediate and low \textit{Q} range. The purpose of our analysis is to put our current results in perspective with the existing reports in the literature.

\begin{figure}
\centering
\includegraphics[scale=0.4]{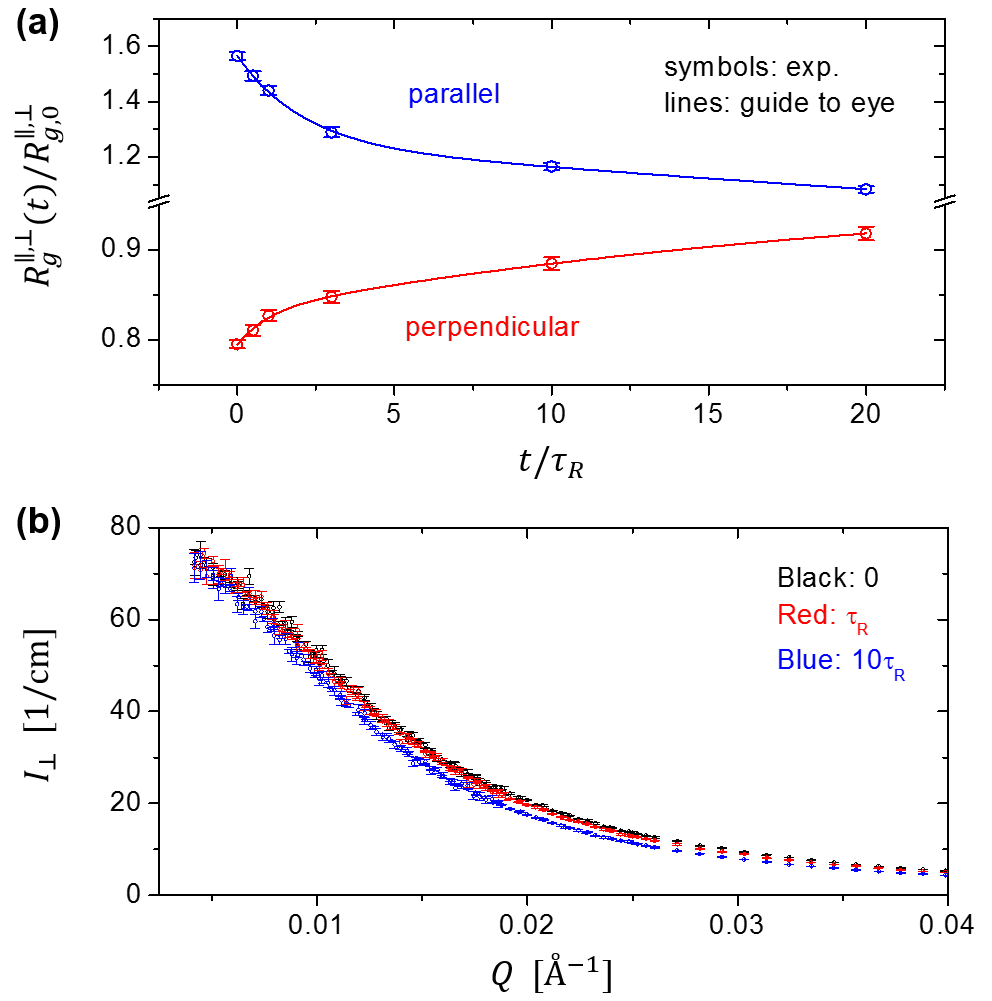}
\caption{(a) Evolution of the radius of gyration in the parallel and perpendicular directions to stretching. $R_{g,0}^{\parallel}$ and $R_{g,0}^{\perp}$ are the equilibrium radius of gyration in the parallel and perpendicular directions, respectively. Obviously, $R_{g,0}^{\parallel}=R_{g,0}^{\perp}$. (b) Evolution of the absolute scattering intensity in the perpendicular direction. The SANS data was collected on the NGB30 SANS beamline at NIST.}
\label{F6}
\end{figure}

Figure \ref{F6}a shows the evolution of radius of gyration during the stress relaxation for both parallel and perpendicular directions. While the tube theory predicts that chain retraction would lead to a non-monotonic change of radius of gyration in the perpendicular direction to stretching (Fig. \ref{F3}b), experimentally, we observe that the $R_g$ in both perpendicular and parallel directions relax \textit{monotonically} towards the equilibrium value. This result is consistent with the findings of Maconnachie \textit{et al}. \cite{Maconnachie} and Bou\'e \textit{et al}. \cite{Boue1, Boue2}, but at odds with the report of Blanchard and coworkers \cite{Blanchard}. To further demonstrate that the fitting by Eq. 15 does faithfully capture the qualitative behavior of the radius of gyration, we present the absolute scattering intensity in the perpendicular direction during the stress relaxation in Fig. \ref{F6}b. We do observe a systematic and monotonic ``shift" of scattering profile towards lower \textit{Q} with increasing relaxation time for \textit{all the data points} we have collected; for the sake of clarity in presentation, only the data for $t=0$, $\tau_R$, and $10\tau_R$ are shown here. Therefore, as long as the fitting procedure is consistently applied over a relatively wide \textit{Q} range, we should only obtain a monotonic trajectory for the $R_g$ in the perpendicular direction.

In this context, we would like to further point out a troubling feature in the paper of Blanchard \textit{et al}. Their Fig. \ref{F2}a shows that in the parallel direction $R_g$ at $0.4\tau_R$ is larger than that at $0.005\tau_R$ ($t\approx 0$). Upon a closer examination, it appears that the difference is slightly greater than the uncertainty represented by the error bars. If so, it would imply that the sample was further stretched during the stress relaxation, which apparently violates the second law of thermodynamics. This puzzling trend suggests the work of Blanchard \textit{et al}. might have some experimental issues, as we shall briefly discuss below, in Section V-E. 

\subsection{Discussion of Possible Explanations}

Since the analyses of the spherical harmonic expansion coefficient $S_2^0(Q)$ and the radii of gyration both reject the characteristic signature of chain retraction, we are now confronted by the inevitable question: what is the explanation for the observed SANS results, if the chain retraction hypothesis does not hold? First, while the tube model is founded on the assumption of affine deformation of the tube \cite{DE2, GraessleyMcLeish}, the idea of non-affine tube deformation has been floating around for quite some time, particularly in the case of crosslinked systems \cite{RubinsteinPanyukov1997, RubinsteinPanyukov2002}. However, incorporating this idea into a dynamic theory of polymeric liquids is still an uncharted territory. Furthermore, the major discrepancy between theory and experiment occurs during the stress relaxation rather than the stress growth, as the chain conformation immediately after the step deformation can be approximated by the affine model (Fig. \ref{F4}c, $t=0$). Therefore, without an alternative mechanism for molecular relaxation, the idea of non-affine deformation alone does not seem to be able to explain the experimental observation.

\begin{figure}
\centering
\includegraphics[scale=0.4]{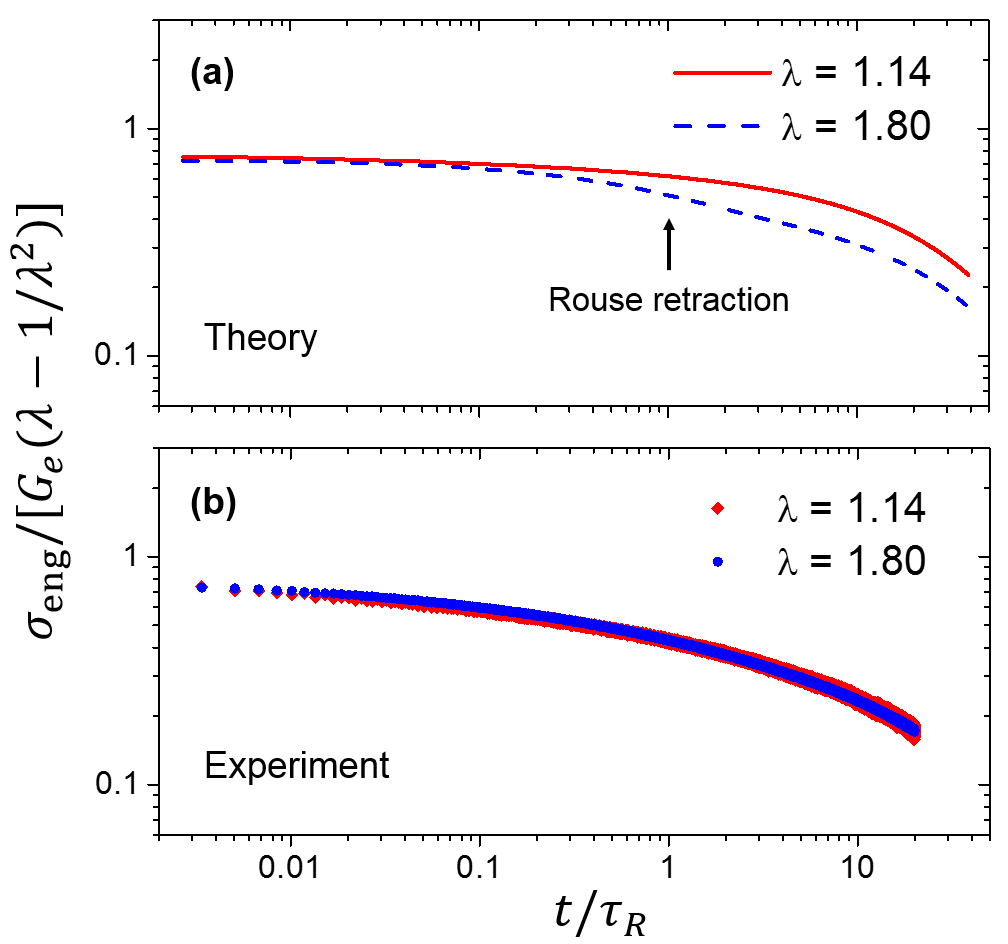}
\caption{Stress relaxation behavior at two different strains $\lambda=1.14$ and $\lambda=1.80$. Here, following the approach of Ref. \cite{ChengSM}, the measured engineering stress $\sigma_{\mathrm{eng}}$ is normalized by $G_e[\lambda-1/\lambda^2]$, where $\lambda$ is the imposed strain. (a) Theoretical predictions from the GLaMM model. (b) Experimental results.}
\label{F7}
\end{figure}

Second, is it possible that the chain retraction does take place, but some other nonlinear effects, unanticipated by the original Doi-Edwards theory, are responsible for the absence of the scattering signature of retraction in SANS? For example, the interplay between test chain motion and topological constraints has been widely recognized \cite{MLD, Graham, SussmanSchweizer1, SussmanSchweizer2, SussmanSchweizer3, SussmanSchweizer4, SussmanSchweizer5}. Could constraint release (CR) during retraction leads to the observed scattering patterns? There is no easy answer to this question. Despite the recent herculean effort by Sussman and Schweizer \cite{SussmanSchweizer1, SussmanSchweizer2, SussmanSchweizer3, SussmanSchweizer4, SussmanSchweizer5} to model the topological constraints in a self-consistent manner, their theory has not yet produced any predictions about SAS behavior for us to compare with our experiments. At this moment, the only available option for us to quantitatively explore the ``constraint release" effect is the GLaMM model, in which the CR can be controlled by tuning the model parameter $c_{\nu}$. Our calculations show, however, that varying $c_{\nu}$ from 0.1 to 1.0 does not change the model prediction of the SANS spectrum for the current step-strain experiment in a substantial way. Additionally, contrary to the prediction of the tube model \cite{SQWangMacro, GrahamComment, ChengSM}, the stress relaxation of our sample (Fig. \ref{F7}) is in agreement with the ``quasi-linear" behavior previously identified by Cheng \textit{et al.} \cite{ChengSM} --- the stress relaxation curves at small strain ($\lambda=1.14$) and large strain ($\lambda=1.80$) can be collapsed by applying an affine scaling to stress. The chain retraction mechanism, on the other hand, would produce a two-step relaxation for $\lambda=1.8$ (Fig. \ref{F7}a). It thus seems unlikely that introducing an additional strong nonlinear CR effect into the tube model would account for the ``quasi-linear" stress relaxation behavior observed experimentally. Our GLaMM calculations indeed confirm that increasing $c_{\nu}$ would result in more pronounced stress drop during relaxation, which is inconsistent with experiment. 

It is interesting to point out a ``nonclassical proposal" that de Gennes \cite{deGennes1991} made many years ago upon learning the SANS investigations by Bou\'e \cite{Boue1, Boue2}. The fact that the signature of retraction was sought but not found in Bou\'e's studies prompted de Gennes to suggest that the chain may indeed not retract at all. However, de Gennes' proposal leads to the prediction that ``there is no strong shear dependence of the viscosity for a monodisperse melt", which clearly contradicts the overwhelming shear-thinning data in the literature.

How about the work by Viovy \cite{Viovy}, who, upon hearing the results of Bou\'e, also proposed his own explanation for the absence of chain retraction in SANS experiments? Viovy's proposal consists of two crucial components: one is the loss of topological constraints on one chain due to the retraction of neighboring chains, and the other is the screening of retraction due to contour length fluctuation. First, as we discussed above, the idea of loss of entanglements at $\lambda=1.8$ does not seem to be compatible with the ``quasi-linear" experimental stress  data, which already lacks the feature of accelerated relaxation due to chain retraction (Fig. \ref{F7}). Second, contour length fluctuation is already incorporated in the calculations with the GLaMM model, but its effect on the qualitative behavior of the SANS spectrum is minimal. Overall, the ideas of Viovy have not been fully developed to yield a complete microscopic model for entangled polymers. This makes it difficult for us to thoroughly evaluate his proposal. In particular, since $R_g$ is not an ideal quantity for comparison of theory and experiments, analysis of $S(\bm{Q})$ through the spherical harmonic expansion technique is the preferred approach. Unfortunately, this has not been done in ref. \cite{Viovy}.

So what could be the explanation for the experimental result (Fig. \ref{F5})? We are currently not in a position to propose our own theory. However, we would like to add a few more comments before leaving this section. First, it should be emphasized that although a stretched chain obviously needs to ``retract" in order to return to its equilibrium state, the concept of ``chain retraction" is a construct of the tube model. It refers specifically to the restoration of the arc length of the primitive chain defined by the ``tube". Therefore, the lack of evidence for chain retraction does not imply that the chains do not relax. On the contrary, both the stress measurements (Fig. \ref{F4}b) and SANS patterns (Fig. \ref{F5}) suggest that the system does continuously relax towards the equilibrium state. Therefore, the issue of ``chain retraction" is about the \textit{pathway} through which the chain relaxes. In other words, it is a matter of the particular molecular relaxation mechanism that an entangled polymer undergoes after a step deformation. Second, the signature patterns of chain retraction in SANS experiments are the consequence of the assumption of ``decoupled" stretch and orientation relaxation in the tube model. The peak shift and anisotropy inversion are directly tied to the physical picture that the contour length equilibrates on the time scale of $\tau_R$, while it takes $\tau$ to completely relax the orientation through reptation. Our analysis with a network-type phenomenological model \cite{Stresspaper} suggests that it is possible to simultaneously describe both the SANS spectrum and stress by assuming \textit{coupled} stretch and orientational dynamics. The details of our quantitative analysis will be published in a future paper.

\subsection{Comments on the Previous Work}

At this point, it seems imperative for us to comment on the previous work of Blanchard \textit{et al.} \cite{Blanchard}, which is the only paper in the literature that claims direct observation of chain retraction by SANS. Their report contradicts not only the current work, but also at least two other independent studies \cite{Maconnachie, Boue1}. Rather than speculating what might have gone wrong in the work of Blanchard \textit{et al.}, we would instead emphasize the steps we have taken to improve the execution of the experiments and data analysis.

First, our stretching experiments were conducted in the Forced Convection Oven (FCO) of the RSA-G2 Solids Analyzer (TA Instruments), which is a well-tested commercial sample environment. We further verified the uniformity and stability of the temperature by monitoring the built-in upper and lower resistance temperature detectors (RTD) of the oven, as well as an additional RTD that we attached to the lower sample clamp.

Second, prior to the final SANS experiments at NIST, we cross-examined the performance of three beamlines (the NGB30 SANS beamline at NIST, the EQ-SANS at SNS, and the D22 at ILL) for 2D data analysis, where every ``pixel" counts. We confirmed that all three beamlines give consistent results for the same quenched samples and ruled out any uncertainty due to the performance of the instrument.

Third, we took care to provide a complete characterization of both the linear and nonlinear viscoelastic properties of the sample. Detailed rheological information was not available in most of the previous investigations on this topic \cite{Maconnachie, Boue1, Blanchard}. This lack of sufficient information on viscoelastic behavior, in our opinion, has hampered discussions of the existing SANS experiments on deformed polymers. The consistency of our stretching experiments on RSA-G2 is demonstrated by the recorded stress response (Fig. \ref{F4}b) --- the stress rise and relaxation data of different runs essentially collapse onto the same envelope.

Fourth, the width and thickness of our quenched samples were carefully measured. We verified that the dimensions of the samples were consistent with the applied macroscopic strain. It is important to note that accurate sample thickness is critical for determination of the absolute scattering intensity. The work of Blanchard \textit{et al}. did not describe how the sample thickness was obtained --- this is a non-trivial issue for their soft, compressible polyisoprene sample. In contrast, the thickness measurement for the high-$T_g$ polystyrene is rather straightforward.

Fifth, we confirmed that our quenched samples has uniform stress distribution by performing birefringence measurements. It appears that none of the previous studies \cite{Maconnachie, Boue1, Blanchard} conducted such a test to verify the qualities of their samples. This was particularly a challenge for Blanchard \textit{et al.}, as their low-$T_g$ sample could only be examined in-situ, \textit{i.e.}, on the SANS beamline.

Sixth, we performed the stretching and relaxation experiments at the same temperature for all the samples. This design avoided the potential complications in the previous studies \cite{Boue1, Blanchard} that utilized the time-temperature superposition principle \cite{Ferry}.

Last, but not least, as we have repeatedly stressed in this article, our spherical harmonic expansion approach makes full use of the entire 2D SANS spectrum, as opposed to the traditional method based on partial information along parallel and perpendicular directions. Moreover, the model-independent nature of the method allows us to circumvent the ambiguity associated with $R_g$ analysis. 

\section{Concluding Remarks and Summary}

In summary, building upon the idea of spherical harmonic decomposition, a new framework has been developed to fingerprint macromolecular deformation from small-angle scattering experiments. The spherical harmonic expansion analysis permits a direct and unambiguous comparison of SANS experiments with the theoretical picture of the tube model. The chain retraction hypothesis of the tube model is not supported by the new SANS measurements of well-entangled polystyrenes after a large step uniaxial extension. Since the tube theory is of paramount importance for our current understanding of the flow and deformation behavior of entangled polymers, the invalidation of the chain retraction hypothesis has immense ramifications. It should be emphasized, however, that the current investigation is only concerned with the tube approach in the \textit{nonlinear} rheological regime. In other words, our work does not question the linear part of the tube theory. Conversely, studies of contour length fluctuations in the equilibrium state \cite{Wischnewski2002} should not be used to infer the validity of chain retraction in the non-equilibrium state. 

Finally, although the application of small-angle scattering in deformed polymers has a long history, the full power of the rheo-SAS technique is yet to be unearthed. The spherical harmonic expansion approach employed in this work is surely not limited to entangled polymeric liquids, but is also applicable to a wide variety of complex fluids and soft solids. The spectrum decomposition method not only provides a convenient way for comparing experimental results with the predictions from statistical and molecular models, but also allows many new questions to be asked, including the affineness, symmetry, and heterogeneity of macromolecular deformation.

\begin{acknowledgements}
This research was sponsored by the Laboratory Directed Research and Development Program of Oak Ridge National Laboratory, managed by UT Battelle, LLC, for the U.S. Department of Energy. W.-R. Chen acknowledges the support by the U.S. Department of Energy, Office of Science, Office of Basic Energy Sciences, Materials Sciences and Engineering Division. J. Liu and Z. Zhao thank the financial support by the NSF Polymer Program (DMR-1105135). The polymer synthesis and characterization were carried out at the Center for Nanophase Materials Sciences, which is a DOE Office of Science User Facility. We acknowledge the support of the National Institute of Standards and Technology, U.S. Department of Commerce, for providing the neutron research facilities used in this work. The statements, findings, conclusions, and recommendations are those of the authors and do not necessarily reflect the view of NIST or the U.S. Department of Commerce. Access to NGB30mSANS was provided by the Center for High Resolution Neutron Scattering, a partnership between the National Institute of Standards and Technology and the National Science Foundation under Agreement No. DMR-1508249. Our preliminary SANS experiments were performed at the EQ-SANS beamline of the Spallation Neutron Source which is a DOE Office of Science User Facility, and the D22 beamline from the Institut Laue-Langevin.
\end{acknowledgements}

\end{document}